\documentclass[%
 aip,
 amsmath,amssymb,
 reprint,%
 groupedaddress
]{revtex4-1}

\usepackage{graphicx}
\usepackage{dcolumn}
\usepackage{bm}

\usepackage[utf8]{inputenc}
\usepackage[T1]{fontenc}
\usepackage{newtxtext}
\usepackage{newtxmath}
\usepackage{etoolbox}

\usepackage{amsmath,amssymb}
\usepackage{graphicx}
\usepackage{gensymb}

\usepackage{tikz}
\usetikzlibrary{shapes.geometric,arrows}

\usepackage{hyperref}
\hypersetup{
colorlinks=true,
citecolor=blue,
urlcolor=blue,
linkcolor=blue
}

\makeatletter
\def\@email#1#2{%
 \endgroup
 \patchcmd{\titleblock@produce}
  {\frontmatter@RRAPformat}
  {\frontmatter@RRAPformat{\produce@RRAP{*#1\href{mailto:#2}{#2}}}\frontmatter@RRAPformat}
  {}{}
}%
\makeatother
\begin{document}

\preprint{AIP/123-QED}

\title{Anisotropic propagation of GHz surface and bulk acoustic waves in gallium arsenide studied by random scattering}
\author{Thomas Steenbergen*}
    \email{steenbergen@physics.leidenuniv.nl}
\author{Maja Wohlfarth}
\author{Pim Veefkind}
\author{Matteo Fisicaro}
\author{Wolfgang L\"offler}

\affiliation{
Huygens-Kamerlingh Onnes Laboratory, Leiden Institute of Physics, Leiden University, The Netherlands
}

\date{\today}

\begin{abstract}
Understanding the complex anisotropic acoustic propagation in crystals is crucial for optimizing the performance of surface and bulk acoustic wave devices. Here, we investigate the anisotropy and coupling of GHz acoustic modes in (001)-cut gallium arsenide through theory and experiment. We first numerically calculate the angle-dependent phase velocities for surface and bulk modes, and we provide a code which can easily be adapted to different material systems. We validate our theoretical model experimentally by exciting surface modes with an interdigital transducer, and achieve omnidirectional acoustic propagation through random scattering of the acoustic waves. We measure the complex acoustic field with a scanning optical interferometer, and extract the angle-dependent velocities of surface and bulk modes using Fourier domain analysis. Our method could be used for the optimization of GHz-range classical and quantum acoustic devices by identifying surface and bulk modes.

\end{abstract}

\maketitle

\section{Introduction}
Anisotropic crystalline materials enable the piezoelectric excitation of high-frequency surface acoustic waves (SAWs), while simultaneously giving rise to complex, direction-dependent acoustic propagation. The small wavelength of GHz-range SAWs compared to their electromagnetic counterpart enables the miniaturization of filters in telecommunications, \cite{Ali2025, Liu2025} while their surface confinement is widely employed in chemical \cite{Li2023} and biological\cite{Zeng2024} sensing applications. More recently, GHz SAWs have emerged as promising carriers of quantum information, due to their long coherence times and their ability to couple to a variety of quantum systems. These include superconducting qubits, \cite{Gustafsson2014, Manenti2017, Moores2018} point defects in diamond \cite{Golter2016} and 2D materials, \cite{Patel2024} and semiconductor quantum dots hosted in gallium arsenide (GaAs).\cite{Wei2018, Decrescent2022, DeCrescent2024} Understanding the complex anisotropic acoustic propagation in crystals is crucial for optimizing the performance of SAW-based (quantum) devices. \cite{Msall2020} 

\begin{figure}[h]
    \centering
    \includegraphics[width=01\linewidth]{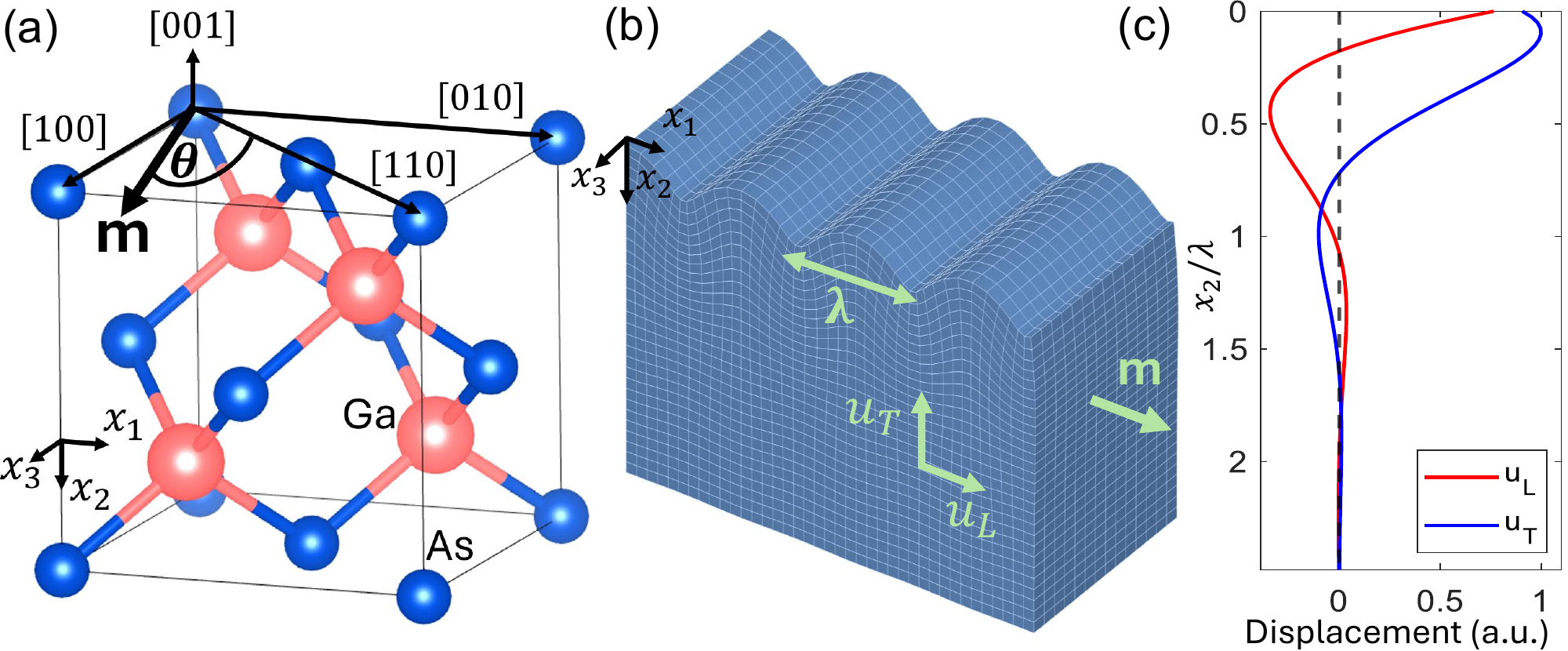}
    \caption{Crystal structure and surface wave displacement. GaAs crystal structure (a), where the coordinate system $(x_1,x_2,x_3)$  is aligned with the crystallographic directions. Sketch of a SAW propagating in the $x_1$ direction (b), where the surface is defined by the $(x_1,x_3)$ plane. Depth-dependent longitudinal ($u_L$) and transverse ($u_T$) displacements (c) for the same SAW mode.}
    \label{fig:Theory_1}
\end{figure}

The anisotropic nature of acoustic propagation in GaAs stems from its zinc-blende crystal structure, which is shown in Fig. \ref{fig:Theory_1}(a). The mechanical polarization of SAWs generally consists of three components, but in high-symmetry directions [100] and [110], only the longitudinal and vertical transverse components remain, as illustrated in Fig. \ref{fig:Theory_1}(b). The SAW displacements decay exponentially into the material, as shown in Fig. \ref{fig:Theory_1}(c). The electromechanical coupling in (001) cut GaAs is relatively weak ($\kappa^2=$ 0.07$\%$),\cite{Royer1996} but is essential for the electrical excitation of acoustic waves using interdigital transducers (IDTs). As the coupling vanishes along the [100] axis and is maximal along the [110] direction, electrical excitation is often performed along the [110] direction.

Calculating the angle-dependent velocities of surface modes in piezoelectric materials is nontrivial, as it requires solving the piezo-elastic wave equations while enforcing surface boundary conditions. Several theoretical approaches have been developed for this purpose. The Stroh formalism,\cite{Stroh1962} which is used here, offers a matrix-based framework in which boundary conditions are easily enforced. Partial-wave methods \cite{Solie1973} offer a more intuitive formulation \cite{Hakoda2018} and are readily extended to layered structures, while Green’s function approaches provide time-domain waveforms, energy flux, and group velocities.\cite{Every1997, Maznev2003} Given the complexity of such calculations, we include ready-to-run code \cite{Steenbergen2026} to facilitate calculations for a variety of crystalline materials. 

Measurements of the angle-dependent SAW phase velocity in GaAs have been reported using Brillouin spectroscopy \cite{Kuok2001} and by varying the orientation of IDTs, \cite{Powlowski2019} while group velocities have been measured using an optical pump–probe technique based on laser-generated SAW pulses. \cite{Maznev2003} Bulk acoustic wave (BAW) velocities have been measured along high-symmetry directions [100] and [110], \cite{Bateman1959, Kuok2000} but full angle-dependent measurements have not been reported to date. 

In this work, we first numerically calculate the angle-dependent velocities of surface and bulk acoustic waves in GaAs. We compare the theoretical results with experimental measurements, where SAWs are generated using an IDT and omnidirectional propagation is obtained by randomly distributed scattering centers. \cite{Laude2011} The out-of-plane amplitude and phase of the acoustic field are mapped using a scanning optical interferometer, \cite{Fisicaro2025} and the angle-dependent velocities are extracted using complex Fourier analysis.\cite{Steenbergen2025} Since the presented measurement technique allows for the simultaneous measurement of surface and bulk modes, it can be used to optimize acoustic devices by identifying these modes.

\section{Acoustic wave theory}
The coupled equations that need to be solved to find the different modes of acoustic propagation in piezoelectric solids combine Hooke’s law and Newton’s second law with piezoelectric coupling: \cite{Royer1996}

\begin{subequations}
\label{eq:CoupledWaveEquations}
    \begin{align}
    c_{ijkl} \frac{\partial^2 u_l}{\partial x_j \partial x_k} + e_{kij} \frac{\partial^2 \Phi}{\partial x_k \partial x_j} &= \rho \frac{\partial^2 u_i}{\partial t^2} \label{eq:MechWaveEquations},\\
    -\epsilon_{jk} \frac{\partial^2 \Phi}{\partial x_j \partial x_k} + e_{jkl} \frac{\partial^2 u_l}{\partial x_j \partial x_k} &= 0 \label{eq:ElecWaveEquation},
\end{align}
\end{subequations}

\noindent where $i,j,k,l = 1,2,3$ and $c_{ijkl}$ is the elastic stiffness tensor, $u_l$ the mechanical displacement, $e_{ijk}$ the piezoelectric tensor, $\Phi$ the electrical potential, $\rho$ the mass density and $\epsilon_{ij}$ the dielectric tensor. The Cartesian coordinate system ($x_1,x_2,x_3$) is aligned with the crystal axes of GaAs as shown in Fig. \ref{fig:Theory_1}(a). All material tensors and properties of GaAs are provided in Appendix A.

\begin{figure}[h]
    \centering
    \includegraphics[width=1\linewidth]{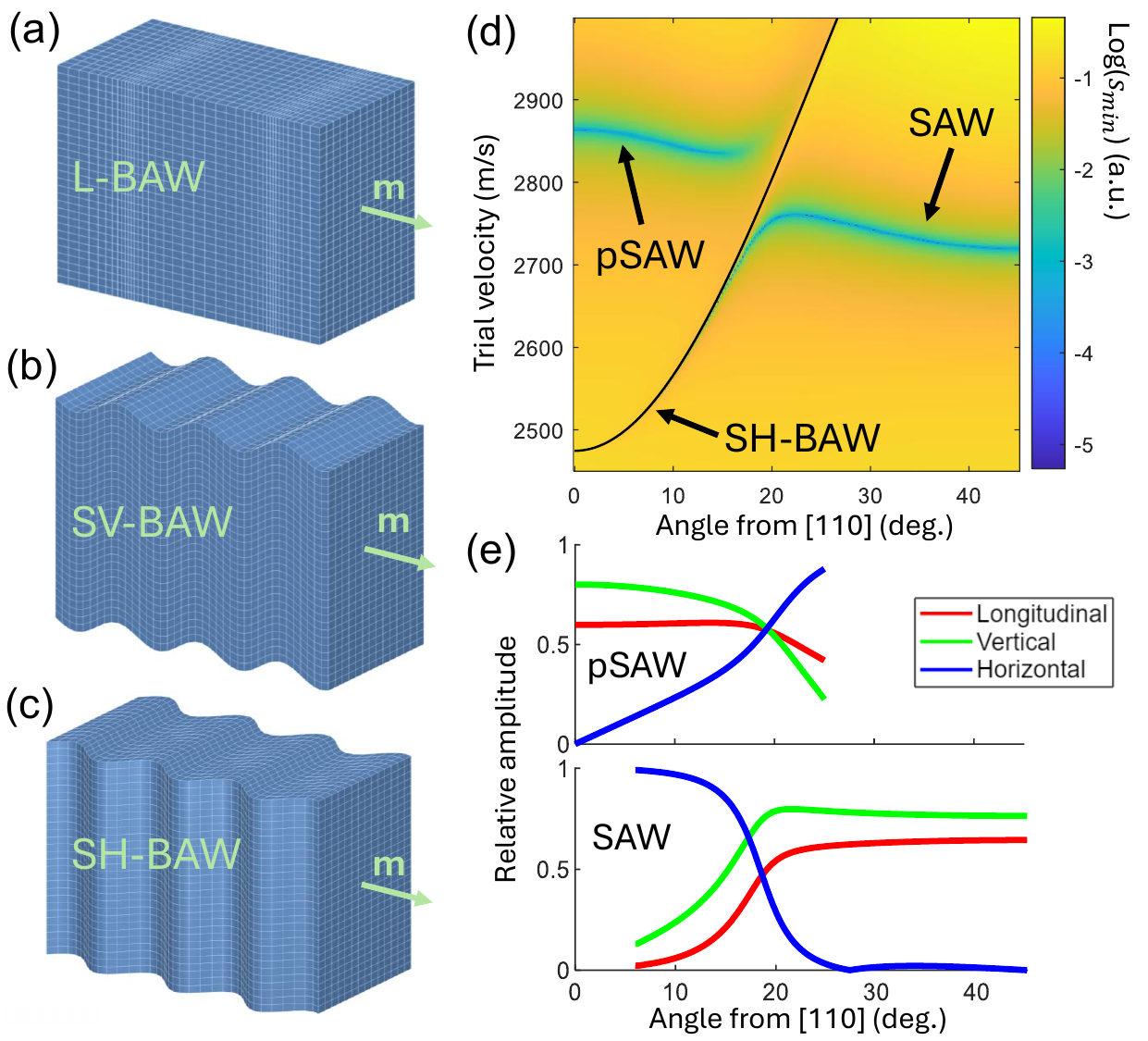}
    \caption{BAW displacements and surface mode velocities and polarization. Sketches (a-c) of the mechanical displacement of the three BAW modes, propagating in the [100] direction. Theoretical boundary condition fullfillment (d), $s_{min}$, of the four chosen partial waves from the Stroh formalism, plotted on a logarithmic scale between the [110] and [100] crystal axes. Relative angle-dependent amplitudes (e) of the mechanical polarizations of the SAW and pSAW modes.}
    \label{fig:Theory_2}
\end{figure}

To find bulk wave solutions, we assume the medium to be infinite in all dimensions, allowing plane wave solutions without boundary conditions. We use the following combined plane wave ansatz for $\mathbf{u}$ and $\Phi$:

\begin{equation}
\begin{bmatrix} \mathbf{u}(\mathbf{x},t) \\ \Phi(\mathbf{x},t) \end{bmatrix}
=
 \mathbf{a}
\exp\Big[i k \big(\mathbf{m} \cdot \mathbf{x} - v t \big)\Big],
\label{eq:BAW_ansatz_v}
\end{equation}

\noindent where $k$ is the wave number, $\mathbf{m}$ the propagation direction vector, $v$ the phase velocity and $\mathbf{a} = [\mathbf{a}_m, a_\Phi]^T$ is the 4D state vector containing the mechanical displacement vector $\mathbf{a}_m$ and the electric potential $a_\Phi$. After substituting the plane-wave ansatz (Eq. (\ref{eq:BAW_ansatz_v})) into Eqs. (\ref{eq:CoupledWaveEquations}a,b) and eliminating the potential $\Phi$, we obtain the Christoffel eigenvalue equation: \cite{Royer1996}

\begin{equation}
\label{eq:Christoffel}
\tilde{\mathbf{\Gamma}} \mathbf{a}_m= \rho  v^2 \mathbf{a}_m,
\end{equation}

\noindent  where $\tilde{\mathbf{\Gamma}}$ is the 3×3 piezoelectric Christoffel matrix, which is constructed from $c_{ijkl}$, $e_{ijk}$, $\epsilon_{jk}$ and the propagation direction $\mathbf{m}$ (see Appendix B). Finding the different modes of propagation and their phase velocities now reduces to solving the eigenvalue problem (Eq. \ref{eq:Christoffel}), where the eigenvectors $\mathbf{a}_m$ describe the mechanical displacement, and the eigenvalues $\rho v^2$ determine the corresponding phase velocities. To obtain angle-dependent velocities, we define the propagation vector $\mathbf{m}=\big(\cos{(\frac{\pi}{4}-\theta)},0, \sin{(\frac{\pi}{4}-\theta)}\big)$ and determine the solutions for the in-plane angle $\theta$, see Fig. \ref{fig:Theory_1}(a). We find 3 modes: a (quasi-) longitudinal mode (L-BAW), a \mbox{(quasi-)} horizontal shear mode (SH-BAW) and a vertical shear mode (SV-BAW). In general, the L-BAW and SH-BAW are coupled and therefore referred to as quasi-modes, but along the [100] and [110] directions they decouple into pure polarizations. Sketches of the mechanical displacements of these bulk modes propagating along the [100] direction are shown in Fig. \ref{fig:Theory_2}(a-c).

Now, to construct surface modes, we first define a half-space where the material surface is spanned by the $(x_1, x_3)$ plane, with inward surface normal $\mathbf{n} = (0,1,0)$ [see Fig. \ref{fig:Theory_1}(b)]. We assume a stress-free, electrically open-circuited material surface and search for surface-bound solutions of the form: \cite{Tanuma2007}

\begin{equation}
\mathbf{U}_n=
\begin{bmatrix} \mathbf{u}(\mathbf{x},t) \\ \Phi(\mathbf{x},t) \end{bmatrix}
=
\mathbf{a}_n \, \exp\Big[ik \big(\mathbf{m}\cdot \mathbf{x} + p_n\, \mathbf{n}\cdot \mathbf{x} - v t \big)\Big],
\label{eq:SAW_ansatz}
\end{equation}

\noindent which represents a partial surface wave (with $\mathbf{a}, k, \mathbf{m}, v$ defined as above), with an additional complex decay parameter $p$, indicating the decay into the substrate along $\mathbf{n}$. A surface mode then, $\mathbf{U}_{SM}$,  consists of a linear combination of 4 partial waves: \cite{Tanuma2007}

\begin{equation}
\mathbf{U}_{SM}= \sum_{n=1}^4 B_n\mathbf{U}_n,
\label{eq:SAW_lin_comb}
\end{equation}

\noindent where $B_n$ are complex coefficients representing the relative weights of the partial waves to the surface mode.

To determine the four partial wave solutions $\mathbf{U}_n$ and to impose the boundary conditions for a range of propagation angles and velocities, we employ the dynamic elastic Stroh formalism, \cite{Tanuma2007} extended to piezoelectric media.\cite{Hwu2008, Hwu2022} The Stroh formalism reformulates the coupled wave equations (Eqs. \ref{eq:CoupledWaveEquations}(a,b)) together with the ansatz (Eq. \ref{eq:SAW_ansatz}) into an 8-dimensional eigenvalue problem:

\begin{equation}
    \mathbf{N}\begin{bmatrix} \mathbf{a} \\ \mathbf{l} \end{bmatrix} = p
    \begin{bmatrix} \mathbf{a} \\ \mathbf{l} \end{bmatrix},
\end{equation}

\noindent where $\mathbf{N}$ is the $8\times8$ Stroh matrix, constructed from $c_{ijkl}$, $e_{ijk}$, $\epsilon_{jk}$, the trial velocity $v$ and the propagation direction $\mathbf{m}$, see Appendix C for its exact construction. The Stroh eigenvector $[\mathbf{a,l}]^{T}$ contains besides the state vector $\mathbf{a}$ the generalized traction vector $\mathbf{l} = [\boldsymbol{\sigma}, D]^T$, which includes the stress vector $\boldsymbol{\sigma}$ and the electric displacement $D$ of the partial wave at the surface. The complex decay parameters $p$ are the eigenvalues of Stroh matrix, and come in conjugate pairs. In the following, we outline the procedure for obtaining the four partial waves $\mathbf{U}_n$, and how the angle-dependent velocity and coefficients $B_n$ can be obtained by imposing boundary conditions. A flowchart of this procedure is provided in Appendix D.

We first construct the Stroh matrix for a range of propagation angles and trial velocities. By calculating its eigenvectors $\mathbf{a}$ and corresponding eigenvalues $p$, we obtain 8 partial wave solutions in the form of Eq. (\ref{eq:SAW_ansatz}). To identify surface-bound modes, we select the four partial waves with the highest positive imaginary parts of $p$, $\Im{(p)}$, corresponding to waves that decay into the substrate. Depending on the propagation angle and the trial velocity, we find either 4 partial waves with $\Im{(p)}>0$ or only three partial waves with $\Im{(p)}>0$ (the other partial wave having $\Im{(p)}=0$). We thus find a mode strongly confined to the surface, the \textit{pure} surface acoustic wave (SAW), and a mode that is more weakly confined to the surface, of which one partial wave does not decay with depth, the \textit{pseudo} surface acoustic wave (pSAW). 

Once the partial waves $\mathbf{U}_n$ have been obtained for each propagation angle and trial velocity, we must now determine the velocity for which the boundary conditions can be satisfied by an appropriate choice of coefficients $B_n$. The boundary information of the four selected partial waves is contained in their generalized traction vectors $\mathbf{l}_n$, from which we construct the 4x4 traction matrix $\mathbf{L} = [\mathbf{l_1},\mathbf{l_2},\mathbf{l_3},\mathbf{l_4}]$. \cite{Tanuma2007} Stress-free and open-circuited boundary conditions ($\sigma_i=0, D=0$) are satisfied if there exists a vector of coefficients $\mathbf{B}=[B_1,B_2,B_3,B_4]^T$ ($B_n$ in Eq. (\ref{eq:SAW_lin_comb})), such that:

\begin{equation}
    \mathbf{L} \mathbf{B}=0,
\end{equation}

\noindent which is the case if $\det(\mathbf{L})=0$. \cite{Tanuma2007} We evaluate this numerically using a singular value decomposition (SVD) of the traction matrix $\mathbf{L}$. If the smallest singular value $s_{\min}$ vanishes, then there is a combination of coefficients $B_n$, such that surface mode $U_{SM}$ satisfies the boundary conditions. We plot $s_{min}$ for a range of angles and velocities in Fig. \ref{fig:Theory_2}(d). We observe two branches where $s_{min}$ is minimized and where the boundary conditions are thus fulfilled. These branches represent the angle-dependent phase velocities of the SAW mode and the pSAW mode. For these modes, the coefficients $B_n$ are obtained from the right singular vector corresponding to $s_{min}$. Together with the four selected eigenvalues $p_n$ and eigenvectors $\mathbf{a_n}$, we now have constructed for both modes the complete depth-dependent mechanical polarization and electrical potential $\mathbf{U}_{SM}$ for every propagation angle. We plot the angle-dependent amplitude of the mechanical polarization at the surface for the pseudo mode (pSAW, upper figure) and the pure mode (SAW, lower figure) in Fig. \ref{fig:Theory_2}(e). Here we find that the mechanical polarization generally comprises three components. Along [110] for the pSAW and along [100] and 27° from [110] for the SAW, the mechanical polarization consists of only vertical and longitudinal components.

We observe in Fig. \ref{fig:Theory_2}(d) that the pure surface mode (SAW) exhibits an avoided crossing with the SH-BAW mode, \cite{Alzina2025,Tarasenko2021} suggesting mode coupling for angles smaller than 20$\degree$ from the [110] axis. Approaching the anti-crossing, we find that the SAW mode acquires a strong horizontal component (Fig. \ref{fig:Theory_2}(e)), which is the main polarization of the SH-BAW. This suggests that the coupling between the SAW and the SH-BAW can indeed explain the anti-crossing. Finally, we note that the non-decaying partial wave of the pseudo mode vanishes precisely in the [110] direction due to the crystal symmetry. As a result, the pSAW mode is here non-leaky. \cite{Maznev2003}

As GaAs is weakly piezoeletric, the piezoelectric effect is often disregarded in calculations; the inclusion of the piezoelectric effect changes the velocity of the pSAW in the [110] direction only by 2.4 m/s. Nevertheless, the piezoelectric effect is essential for the calculation of the angle-dependent electromechanical coupling coefficient $\kappa$ for surface modes. This coefficient can be determined from the velocity difference $\Delta v$ between short-circuited ($a_{\Phi}$ = 0) and open-circuit ($D=0$) electrical boundary conditions: $\kappa^2 = 2\frac{\Delta v}{v}$. \cite{Oliner1978}

\section{Experimental setup}
\begin{figure}[t]
    \centering
    \includegraphics[width=1\linewidth]{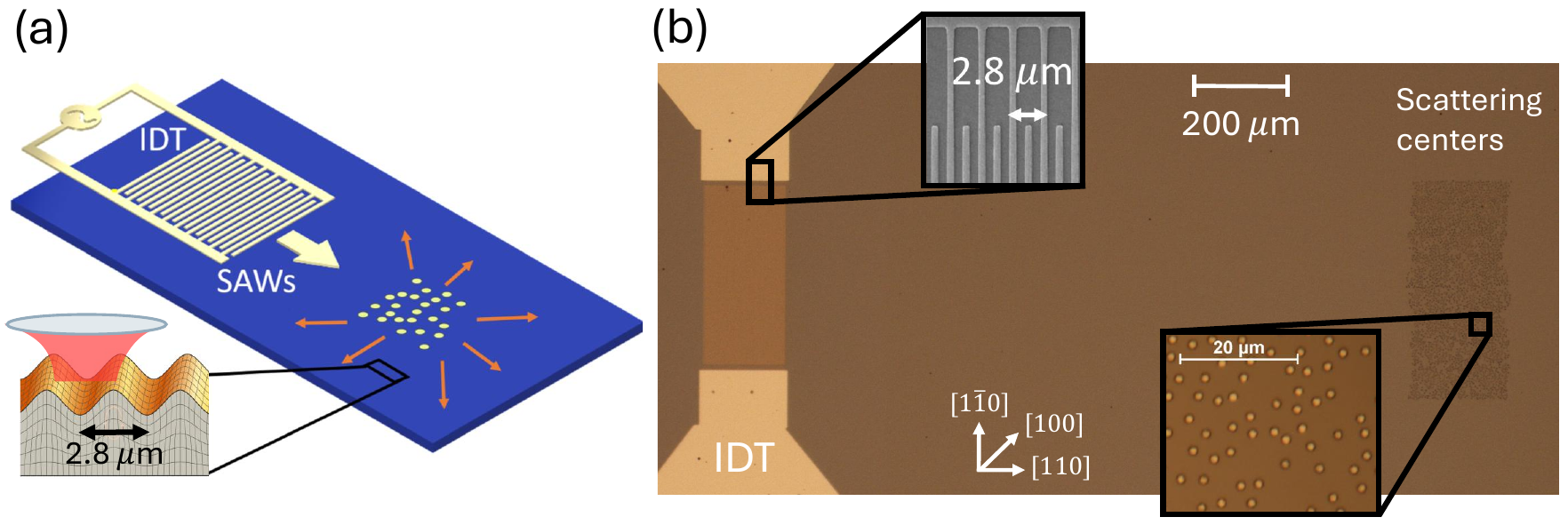}
    \caption{Experimental setup. Schematic (a) of the acoustic scattering sample, where the inset depicts the optical measurement of the acoustic field. Optical microscope image (b) of the acoustic scattering device, with inset showing a scanning electron microscope image of the IDT finger structure.}
    \label{fig:Exp_setup}
\end{figure}

In the experiment, we generate surface modes with an IDT and omnidirectional propagation is achieved by randomly positioned scattering centers, see Fig. \ref{fig:Exp_setup}. The IDT and the disk-shaped scattering centers (diameter 1.4 $\mu$m) are fabricated on a (001)-cut GaAs substrate using an electron-beam lithography lift-off process and consist of a 5 nm titanium adhesion layer, a 40 nm aluminum layer and a 5 nm titanium capping layer. The IDT has 50 finger pairs and the fingers are 700 nm wide and 312 $\mu$m long. The periodicity of the IDT fingers is chosen at 2.8 $\mu$m [see inset Fig. \ref{fig:Exp_setup}(b)], to efficiently generate surface modes at 1.032 GHz ($v_{pSAW} \approx 2860$ m/s in the [110] direction). 

\begin{figure}[b]
    \centering
    \includegraphics[width=0.85\linewidth]{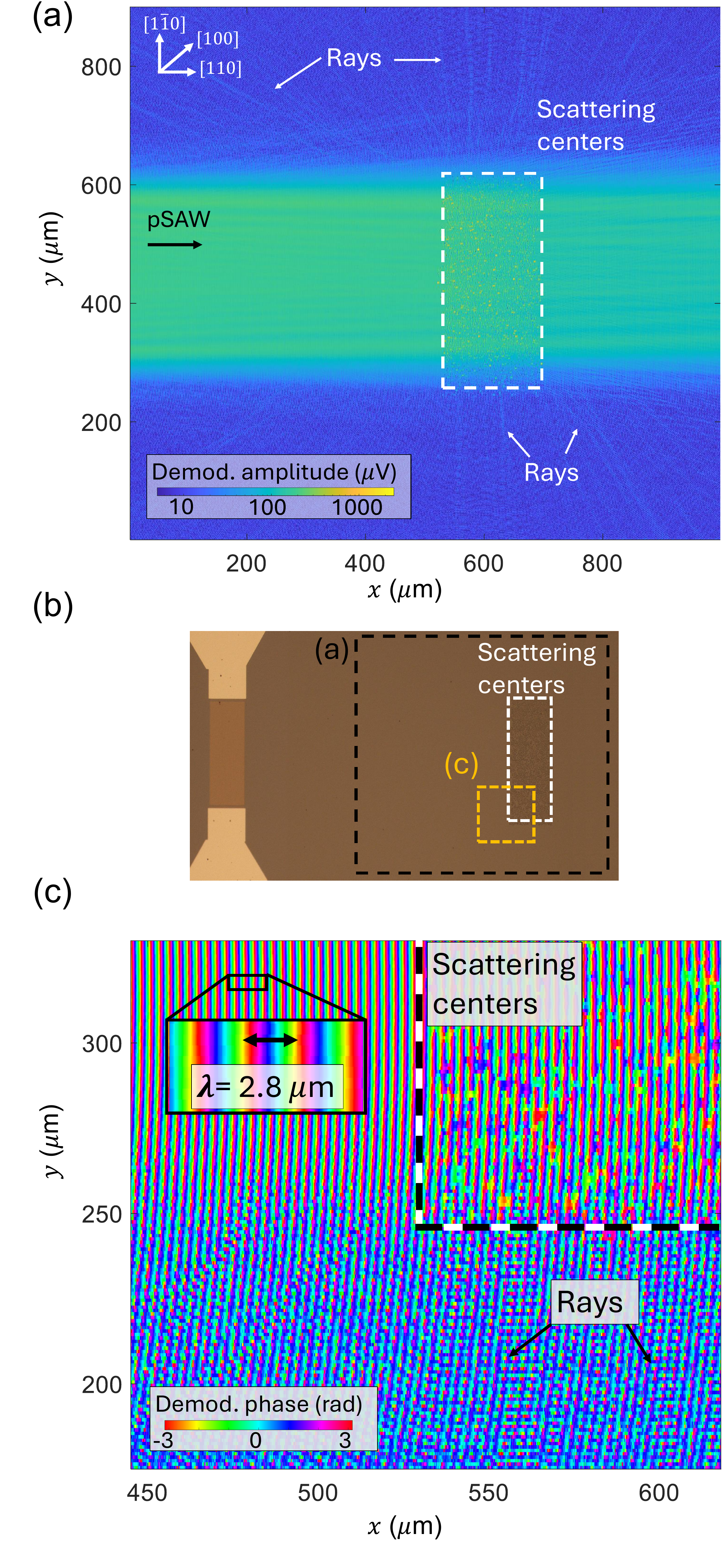}
    \caption{Real space experimental data. Rms amplitude amplitude (a, logarithmic scale) and phase (c) of the 1.03205 GHz demodulated interferometric signal. The measurements were performed in the areas indicated by the black dashed box for (a) and the yellow dashed box for (c) in the microscope image (b).}
    \label{fig:Results_1}
\end{figure}

\begin{figure*}[!t]
    \centering
    \includegraphics[width=.95\linewidth]{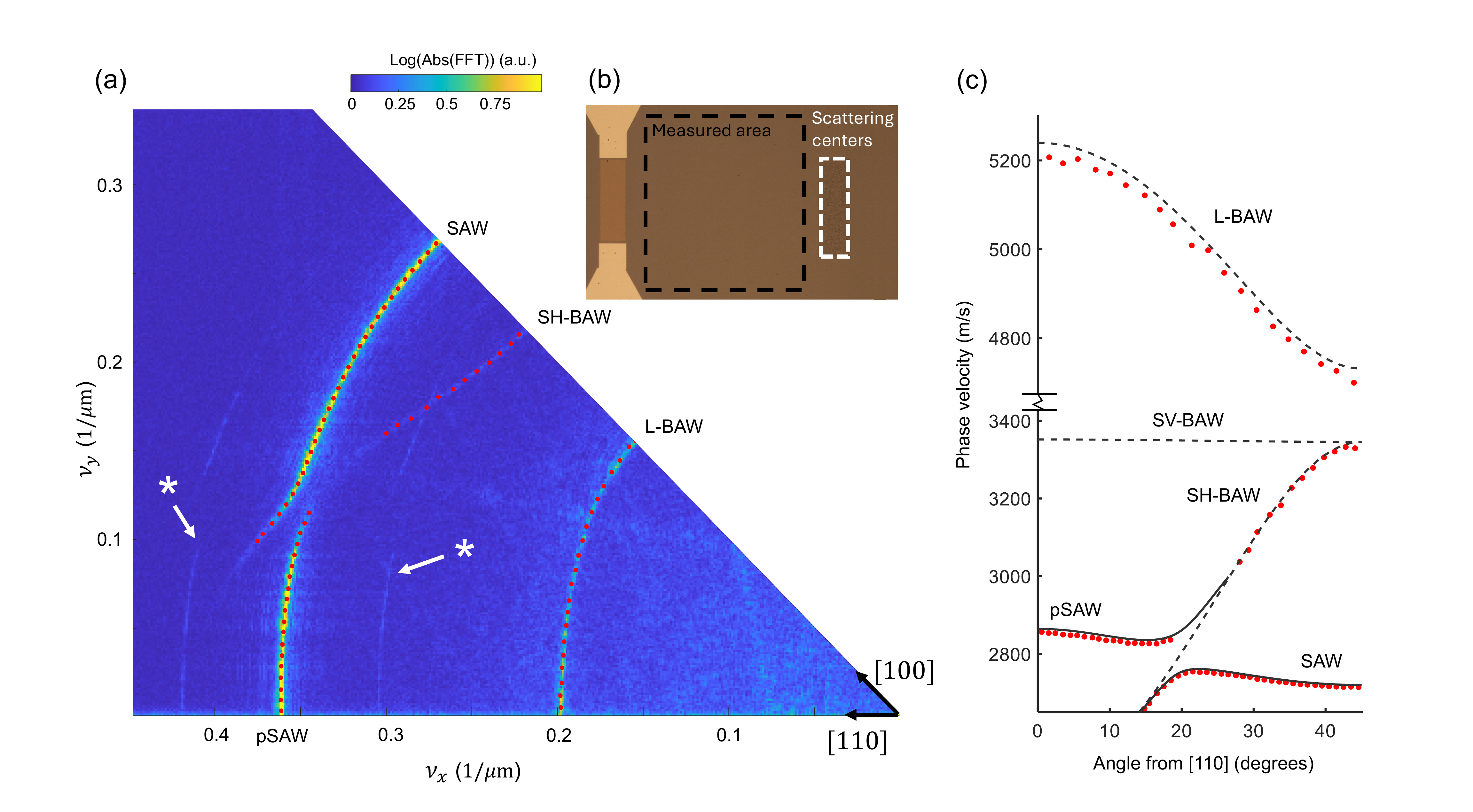}
    \caption{Fourier analysis for acoustic mode identification and comparison to theory. Absolute value (a) of the spatial Fourier transform of the complex measurement of the area indicated by the dashed black box in (b). The data is folded to a 45 degree sector and the values $(\bar{\nu}_x,\bar{\nu}_y)$ are shown in red. Comparison (c) between the measured (red dots) and theoretical angle-dependent velocities of the surface (black curves) and bulk (dashed black curves) acoustic modes.}
    \label{fig:Results_2}
\end{figure*}

The out-of-plane acoustic displacement is measured with an optical scanning Michelson interferometer, \cite{Fisicaro2025} in which laser light is focused on the acoustic scattering sample with a spot size of around 3 $\mu$m [see inset Fig. \ref{fig:Exp_setup}(a)]. The optical path length is modified by the out-of-plane component of the acoustic displacement, which leads to a GHz modulation of the interferometric signal that is read-out using a lock-in technique, allowing for the extraction of the amplitude and phase of the acoustic displacement. The relatively large laser spot size compared to the SAW wavelength reduces the sensitivity, but still provides sufficient signal.\cite{Fisicaro2025} To obtain high-resolution spatial scans, the acoustic sample is mounted on a motorized XYZ translation stage with a step size of 14 nm in the $x$ direction (the [110] crystal axis) and 1 $\mu$m in the $y$ direction (the [1$\overline{1}$0] crystal axis).

\section{Results and discussion}
We show the measurement of the amplitude of the demodulated interferometric signal in Fig. \ref{fig:Results_1}(a), where the measured area is indicated by the dashed black box in Fig. \ref{fig:Results_1}(b). Here, we observe a beam of surface modes generated by the IDT directed towards the scattering area, indicated by the white dashed box. A significant portion of the acoustic field, however, is not affected by the scattering area (around 80\%). Nevertheless, as we will show later, the scattered waves provide sufficient signal to extract the angle-dependent velocities of the acoustic modes in all directions. We also observe rays propagating in a variety of directions from the scattering area, which are due to random-wave interference. A zoomed-in (yellow dashed box in Fig. \ref{fig:Results_1}(b)) phase measurement is shown in Fig. \ref{fig:Results_1}(c). Again, we observe the incoming beam, now oscillating between -$\pi$ and +$\pi$ with the expected periodicity of the pSAW wavelength of around 2.8 $\mu$m, as shown in the inset. In the scattering area, indicated by the dashed black and white lines, we observe that the scatterers affect the plane wave, which results in random patterns of the measured phase. The rays due to random interference are also visible in the phase measurement.  

To distinguish the different acoustic modes by their angle-dependent wavelengths, we perform spatial Fourier analysis. For this, we first combine the amplitude and phase measurements of the area depicted by the dashed black box in Fig. \ref{fig:Results_2}(b) in a complex field, of which we take the spatial 2D fast Fourier transform (FFT). We choose this area to exclude the field in the scattering area, as its random spatial periodicities result in noise and artefacts in the spatial FFT. We select only the back-scattered field (i.e. waves moving to the left in Fig. \ref{fig:Results_1}(a, c)), which corresponds to spatial frequencies $\nu_x>0$. Next, we reduce the remaining 180$\degree$ range of propagation directions to a 45$\degree$ sector by exploiting the zinc-blende crystal symmetry, under which [100] is equivalent to [010], and [110] is equivalent to [1$\bar{1}$0]. First, we fold the data along the $\nu_x$ axis (the [110] crystal axis) by reflecting the $\nu_y<0$ plane onto $\nu_y>0$ plane and averaging of the corresponding FFT amplitudes, resulting in a 90$\degree$ sector. Then, we fold the data in the same way along the $\nu_x=\nu_y$ (the [100] crystal axis), reducing the angular range further to 45$\degree$.

In Fig. \ref{fig:Results_2}(a), we plot the magnitude of the folded FFT between the [110] and the [100] crystal axes. In this figure, we identify acoustic modes by their angle-dependent spatial periodicity $(\nu_x, \nu_y)$. We find a strong signal associated with the pure SAW mode at the [100] axis, which does not fully extend to the [110] axis due to the avoided crossing with the SH-BAW mode. The pSAW mode extends further to the [110] axis. This observation is in line with the theoretical results (see Fig. \ref{fig:Theory_2}(d)). We also observe a clear signal corresponding to the L-BAW and a weaker signal associated with the SH-BAW. We further observe two sidebands (indicated by *) of the surface modes. Since they are equally spaced from the main signal in the $\nu_x$ direction and decrease in amplitude for higher $\nu_x$, we hypothesize that these sidebands are the result of an amplitude modulation in the $x$ direction caused by imperfections of the translation stage. \footnote{We investigated this effect by rotating the sample and found that the sidebands remained dominant along the same axis of the translation stage.}

Finally, we calculate for each acoustic mode the weighted averages $(\bar{\nu}_x,\bar{\nu}_y)$  for a set of angles, which are depicted as red dots in Fig. \ref{fig:Results_2}(a). This is achieved by first performing a Frangi ridge detection algorithm, \footnote{We used the MATLAB \href{https://www.mathworks.com/help/images/ref/fibermetric.html}{\texttt{fibermetric}} function.} which identifies the data points that form the branches in the FFT. Then, we cluster the data points from the ridge detection per acoustic mode using a DBSCAN algorithm. \footnote{We applied the \href{https://www.mathworks.com/help/stats/dbscan-clustering.html}{\texttt{dbscan}} function from the MATLAB Statistics and Machine Learning Toolbox\textsuperscript{\texttrademark}.} Finally, we calculate for each mode the average $(\nu_x,\nu_y)$-position of the selected data points within an angular bin with the FFT amplitude as weight, and we obtain $(\bar{\nu}_x,\bar{\nu}_y)$. We then calculate the propagation angle $\theta$ (from the [110] axis) and the corresponding phase velocity $v(\theta)$ as follows:

\begin{align}
    \theta &= \arctan\!\left(\frac{\bar{\nu}_y}{\bar{\nu}_x}\right), \\
    v(\theta) &= \frac{f}{\sqrt{\bar{\nu}_x^{\,2} + \bar{\nu}_y^{\,2}}},
\end{align}

\noindent where the $f$ is the excitation frequency of 1.03205 GHz. The resulting angle-dependent velocities are shown in Fig. \ref{fig:Results_2}(c), where the red dots are the experimental data and the curves are the theoretical predictions. We find a very good agreement between experiment and theory. 

At first glance, the measurement of BAW modes is surprising, as they lack an out-of-plane displacement component (see Fig. \ref{fig:Theory_2}(a, c)) and therefore should not modulate the optical path length -- the principle underlying our detection technique. However, stress-free boundary conditions alter the displacement profile of the modes near the surface, possibly introducing an out-of-plane component. Such bulk modes satisfying the stress-free boundary conditions at the surface are referred to as surface-skimming or lateral BAWs.\cite{Lewis1977, Dewhurst1986, Sathish2004} In addition, a modulation of the refractive index by the acousto-optic effect \cite{Chaudhary2025, Debye1932} could contribute to the measured signal. Interestingly, despite exhibiting an out-of-plane displacement component, the SV-BAW is not observed and the origin of its absence remains unclear.

\section{Conclusion and outlook}
In conclusion, we measured the angle-dependent velocity of surface and bulk acoustic waves in (001) cut GaAs and we found excellent agreement with theoretical calculations. For our calculations, we employed the piezoelectric Christoffel formalism for bulk modes and the extended Stroh formalism for surface modes, and our published code can easily be extended to different material systems. In the experiment, we generated acoustic waves using an IDT, and omnidirectional propagation is obtained by randomly positioned scattering centers. The out-of-plane acoustic displacement is mapped with a scanning optical interferometer and using complex Fourier analysis, the angular dispersion of surface and bulk modes was obtained. Interestingly, bulk modes, which in the bulk have no out-of-plane displacement, were measured at the surface. This measurement technique, combined with Fourier analysis, could be used for the optimization of (quantum) acoustic devices by identifying surface and bulk acoustic waves. The acoustic scattering device can furthermore be employed to investigate acoustic wave propagation in random media. \cite{Wiersma1997, Vellekoop2007, Hu2008}

\section*{Acknowledgments}
We thank Harry Visser and Peter van Veldhuizen for their help with RF electronics. We acknowledge funding from NWO (680.92.18.04), NWO/OCW (Quantum Software Consortium Nos. 024.003.037 and 024.003.037/3368, Quantum Limits No. SUMMIT.1.1016), and from the Dutch Ministry of Economic Affairs (Quantum Delta NL).

\section*{Author declarations}

\subsection*{Conflict of Interest}
The authors have no conflicts to disclose.

\subsection*{Author Contributions}
\textbf{Thomas Steenbergen:} Conceptualization; Data curation; Investigation; Methodology; Formal Analysis; Visualization;  Writing – original draft; Writing – review \& editing. \textbf{Maja Wohlfarth:} Formal analysis; Investigation; Software. \textbf{Pim Veefkind:} Methodology. \textbf{Matteo Fisicaro:} Conceptualization, Methodology; Writing – review \& editing. \textbf{Wolfgang L\"offler:} Conceptualization; Funding acquisition; Supervision;  Writing – review \& editing.

\section*{Data availability}
The theoretical model and data is online available. \cite{Steenbergen2026} The experimental data underlying the results presented in this paper can be obtained from the authors upon reasonable request.

\appendix
\renewcommand{\theequation}{A.\arabic{equation}}
\setcounter{equation}{0}

\section*{Appendix A: GaAs material properties}
We provide here the material properties of GaAs. The fourth-rank stiffness tensor $c_{ijkl}$ is represented in contracted Voigt notation as the $6\times6$ matrix $c_{\alpha\beta}$ and the third-rank piezoelectric tensor $e_{ijk}$ as the $3\times6$ matrix $e_{i\alpha}$. The tensor indices are mapped to matrix indices according to:
\[
\begin{array}{ccc}
(ij, kl) & \rightarrow & (\alpha, \beta) \\[4pt]
(11) & \rightarrow & 1 \\
(22) & \rightarrow & 2 \\
(33) & \rightarrow & 3 \\
(23), (32) & \rightarrow & 4 \\
(13), (31) & \rightarrow & 5 \\
(12), (21) & \rightarrow & 6 \\
\end{array}
\]

\noindent The elastic tensor for GaAs is given by:
\begin{equation}
\mathbf{c_{\alpha\beta}} =
\begin{bmatrix}
c_{11} & c_{12} & c_{12} & 0     & 0     & 0 \\
c_{12} & c_{11} & c_{12} & 0     & 0     & 0 \\
c_{12} & c_{12} & c_{11} & 0     & 0     & 0 \\
0      & 0      & 0      & c_{44} & 0     & 0 \\
0      & 0      & 0      & 0      & c_{44} & 0 \\
0      & 0      & 0      & 0      & 0      & c_{44}
\end{bmatrix},
\label{eq:A_cij}
\end{equation}

\noindent with: \cite{Royer1996}
\begin{equation*}
c_{11} = 118.8~\mathrm{GPa}, \quad
c_{12} = 53.8~\mathrm{GPa}, \quad
c_{44} = 59.4~\mathrm{GPa}.
\end{equation*}

\noindent The piezoelectric tensor is given by:
\begin{equation}
\mathbf{e}_{i\alpha} =
\begin{bmatrix}
0 & 0 & 0 & e_{14}     & 0 & 0 \\
0 & 0 & 0 & 0 & e_{14}     & 0 \\
0 & 0 & 0 & 0     & 0     & e_{14}
\end{bmatrix},
\label{eq:A_eij}
\end{equation}

\noindent with: \cite{Royer1996}

\begin{equation*}
e_{14} = -0.16~\mathrm{C/m^2}.
\label{eq:A_evalue}
\end{equation*}

\noindent The dielectric tensor of GaAs is given by:
\begin{equation}
\boldsymbol{\epsilon}_{ij} =
\epsilon_{s}
\begin{bmatrix}
1 & 0 & 0 \\
0 & 1 & 0 \\
0 & 0 & 1
\end{bmatrix},
\label{eq:A_eps}
\end{equation}
with: \cite{Royer1996}

\begin{equation*}
\epsilon_s = 12.9\,\epsilon_0, \quad
\epsilon_0 = 8.854\times10^{-12}~\mathrm{F/m}.
\label{eq:A_epsvalue}
\end{equation*}

Finally, the mass density of GaAs is given by $\rho = 5307\,\mathrm{kg\,m^{-3}}$. \cite{Royer1996}

\section*{Appendix B: Christoffel matrix for piezoelectric solids}
Here we show the construction of the $3\times3$ Christoffel matrix $\tilde{\mathbf{\Gamma}}_{il}$ from the propagation vector $\mathbf{m}=(m_1,m_2,m_3)$ and the material tensors. 
The piezoelectric Christoffel matrix elements are given by: \cite{Royer1996}

\begin{equation}
    \tilde{\Gamma}_{il} = \Gamma_{il}+\frac{\gamma_i\gamma_l}{\epsilon},
\end{equation}

\noindent where: 
\begin{subequations}
\label{eq:ChristoffelBlocks}
\begin{align}
\Gamma_{il} &= c_{ijkl} \, m_j m_k, \\
\gamma_i &= e_{kij} m_jm_k,\\
\epsilon &= \epsilon_{jk} m_jm_k,& i,j,k,l &= 1,2,3.
\end{align}
\end{subequations}

\section*{Appendix C: Stroh matrix for piezoelectric solids}
In this Appendix, we outline the construction of the Stroh matrix for anisotropic piezoelectric solids. We extend the dynamic $6\times6$ Stroh matrix for elastic solids \cite{Tanuma2007} to piezoelectric materials by incorporating piezoelectric terms in an \textit{effective} stiffness tensor $\tilde{C}_{ijkl}$, \cite{Hwu2008, Hwu2022} resulting in an $8\times8$ Stroh matrix:

\begin{equation}
\mathbf{N} =
\begin{bmatrix}
\mathbf{N}_1 & \mathbf{N}_2 \\[1mm]
\mathbf{N}_3 & \mathbf{N}_4
\end{bmatrix},
\end{equation}

\noindent where the $4\times4$ sub-blocks are given by:
\begin{equation}
\begin{aligned}
\mathbf{N}_1 &= -\mathbf{T}^{-1}\mathbf{R}^T, \\
\mathbf{N}_2 &= \mathbf{T}^{-1}, \\
\mathbf{N}_3 &= \mathbf{R}\mathbf{T}^{-1}\mathbf{R}^T - \mathbf{Q}, \\
\mathbf{N}_4 &= -\mathbf{R}\mathbf{T}^{-1}.
\end{aligned}
\end{equation}

\noindent The matrices $\mathbf{Q}, \mathbf{R}, \mathbf{T}$ are defined in terms of $\tilde{C}_{ijkl}$ as: \cite{Tanuma2007}
\begin{equation}
\begin{aligned}
Q_{ik} &= \tilde{C}_{ijkl}m_jm_l,  \\
R_{ik} &= \tilde{C}_{ijkl}m_jn_l, \quad j,l = 1,2,3,\\
T_{ik} &= \tilde{C}_{ijkl}n_jn_l, \quad i,k=1,2,3,4.
\end{aligned}
\end{equation}

Up until now, we have constructed the static extended Stroh matrix. We obtain the dynamic Stroh matrix by including a trial velocity $v$ in the first 3 diagonal elements of $Q_{ik}$: \cite{Tanuma2007}

\begin{equation}
Q_{ik} = Q_{ik} - \rho v^2 \delta_{ik} \quad i,k = 1,2,3,
\end{equation}

\noindent where $\rho$ is the mass density of GaAs. The effective tensor $\tilde{C}_{ijkl}$ is obtained by extending the conventional elastic stiffness tensor to include piezoelectric and dielectric couplings: \cite{Hwu2022}
\begin{equation}
\begin{aligned}
\tilde{C}_{ijkl} &= c_{ijkl}, \\    
\tilde{C}_{ij4l} &= e_{lij}, \\
\tilde{C}_{4jkl} &= e_{jkl}, \\
\tilde{C}_{4j4l} &= -\epsilon_{jl}, &   i,j,k,l &= 1,2,3.
\end{aligned}
\end{equation}

\section*{Appendix D: Flowchart of surface mode calculations}
Below we provide a flowchart visualizing the algorithm used for the construction of SAW and pSAW modes and the calculation of their angle-dependent velocities. 

\begin{figure}[h]
    \includegraphics[width=0.5\textwidth]{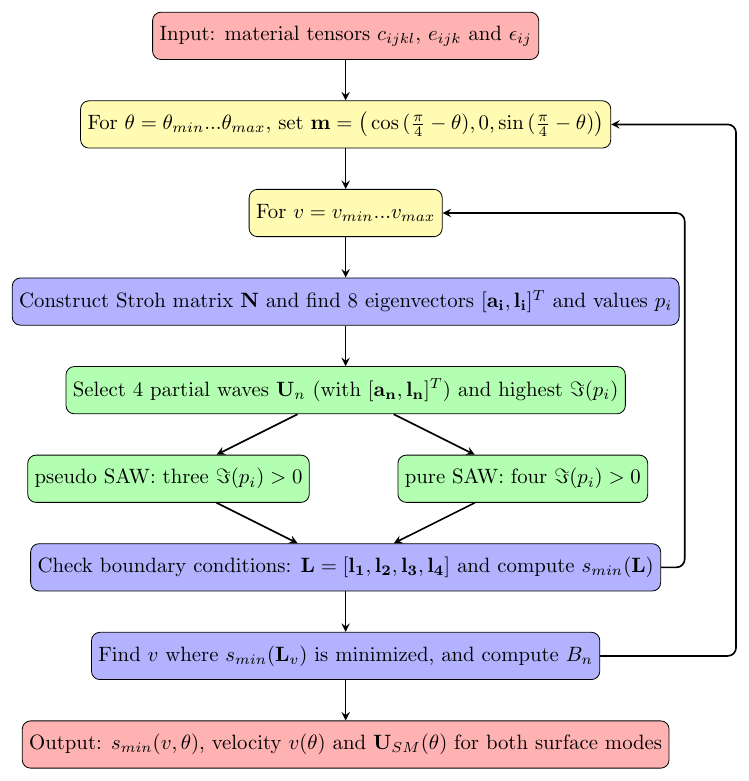}
    \caption{Flowchart of the numerical procedure used to construct SAW and pSAW modes and to determine their angle-dependent velocities.}
\end{figure}

\section*{References}
\bibliography{export}

\end{document}